\documentclass[aps,pra,twocolumn,superscriptaddress]{revtex4-2}

\usepackage[utf8]{inputenc}

\usepackage[titletoc,toc,title,page]{appendix}

\usepackage{csquotes,graphicx}
\usepackage[italian,english]{babel}

\usepackage{microtype} 

\usepackage{bm} 

\usepackage{dsfont} 
\usepackage{amsmath,amssymb,amsthm,thmtools}
\usepackage{mathtools}
\usepackage{cases}
\usepackage{calc,palatino}
\usepackage{mathrsfs} 
\usepackage[normalem]{ulem} 

\usepackage{parskip}
\usepackage{nameref}
\usepackage[colorlinks=true]{hyperref}
\usepackage[nameinlink]{cleveref}
\crefname{appsec}{Appendix}{Appendices}
\crefname{box}{Box}{Box}
\hypersetup{
  colorlinks   = true, 
  urlcolor     = green!80!black, 
  linkcolor    = blue, 
  citecolor    = red!80!black 
}

\usepackage{physics} 

\usepackage{float} 

\usepackage{graphicx}

\usepackage[usenames,dvipsnames,table]{xcolor}
\usepackage{easyReview}

\usepackage{tikz}
\usetikzlibrary{calc,shapes.geometric}

\usepackage{placeins}
\usepackage{multirow,tabularx,booktabs}
\usepackage[most]{tcolorbox}
\newtcbtheorem{tbox}{Box}{enhanced, float*=t, width=\textwidth, label type=box}{box}

\usepackage[printonlyused,withpage,nohyperlinks,smaller]{acronym}

\graphicspath{{./figures/}}

\newcommand{\calD}{\mathcal{D}}

\newcommand{\calL}{\mathcal{L}}

\newcommand{\calZ}{\mathcal{Z}}

\newcommand{\TPM}{{\rm TPM}}
\newcommand{\EPM}{{\rm EPM}}

\begin{document}

\title{Coherent heat exchange in a prethermalizing open quantum system}
\author{Simone Artini}
\let\comma,
\affiliation{Universit\`a degli Studi di Palermo\comma{} Dipartimento di Fisica e Chimica - Emilio Segr\`e\comma{} via Archirafi 36\comma{} I-90123 Palermo\comma{} Italy}
\author{Mauro Paternostro}
\let\comma,
\affiliation{Universit\`a degli Studi di Palermo\comma{} Dipartimento di Fisica e Chimica - Emilio Segr\`e\comma{} via Archirafi 36\comma{} I-90123 Palermo\comma{} Italy}
\affiliation{Centre for Quantum Materials and Technologies\comma{} School of Mathematics and Physics\comma{} Queen’s University Belfast\comma{} BT7 1NN\comma{} United Kingdom}

\author{Salvatore Lorenzo}
\let\comma,
\affiliation{Universit\`a degli Studi di Palermo\comma{} Dipartimento di Fisica e Chimica - Emilio Segr\`e\comma{} via Archirafi 36\comma{} I-90123 Palermo\comma{} Italy}


\begin{abstract}
\noindent
We investigate a simple model exhibiting a prethermal phase, i.e. a metastable state that emerges before full thermalization, through the framework of quantum stochastic thermodynamics. We explore the effects of quantum coherence in the energy eigenbasis of the initial state of the system on the process of heat exchange with a bath, and their contribution to entropy production as quantified by a heat-exchange fluctuation theorem. Such relation is derived using the End-Point Measurement (EPM) scheme, a protocol that accounts for initial quantum coherence in the statistics of energy exchanges resulting from a non-equilibrium process.  
We compare these results with those obtained from the widely used Two-Point Measurement (TPM) scheme which, by construction, fails to capture such quantum effects.

\end{abstract}

\maketitle

\section{Introduction}
In the last few years the degree of control over quantum systems has rapidly increased and, with it, the understanding of quantum dynamics as a whole, from many body systems to open quantum systems. Of particular interest from both a practical and theoretical perspective is the development of a non-equilibrium thermodynamic theory in the quantum regime that is able to describe the energy exchange with the environment of such systems and the onset of irreversibility from the underlying unitary dynamics.

Stochastic thermodynamics has offered the best tools for understanding such properties in mesoscale systems in the classical regime, with the central results being the celebrated Fluctuation theorems (FTs) \cite{crooks1999entropy,jarzynski1997nonequilibrium,jarzynski2004classical}. These relations are able to capture out-of-equilibrium thermodynamic properties and to link them to equilibrium ones in very general settings and are able to explain the onset of irreversibility as a violation of a time-reversal symmetry at the level of the stochastic trajectories of the system. This promising framework has been adopted at the quantum level as well, leading to quantum versions of the FTs that resemble the classical ones \cite{esposito2009nonequilibrium} and that have been experimentally verified on numerous experimental platforms \cite{an2015experimental, cimini2020experimental, zhang2018experimental, batalhao2014experimental}. The protocol adopted, the Two-point measurement scheme, requires two strong energy measurements on the system, one at the initial and one at the final time. However, this approach returns the wrong marginals of initial and final energy distributions and is thus insensitive to \textit{initial} quantum coherences in the energy basis and general quantum correlations that could play a role at the level of energy exchange and entropy production. To overcome this limitation and to give a framework able to capture all possible quantum features, many other protocols have been proposed leveraging, for example, quasiprobability distributions~\cite{lostaglio2022kirkwood, diaz2020quantum} or Dynamic Bayesan Networks \cite{micadei2020quantum} in order to recover the correct energy marginals, which have led to modified versions of the classical FTs.

In order to characterize possible deviations from the usual FTs due to initial coherences in the system and possibly being able to experimentally verify them, it is necessary to look at dynamics that, at least for a large enough time window, preserve information on the initial state. Such a behavior can be found in prethermalizing systems. These are quantum quasi-integrable systems that reach a metastable state before fully thermalizing and are usually associated to a quench in a many body Hamiltonian that breaks its symmetry to a certain degree while keeping intact some of the original integrals of motion \cite{Langen2016,Bertini2015,Kollar2011,angles2020prethermalization,marcuzzi2013prethermalization, marcuzzi2016prethermalization}. Depending on the model under consideration, in this transient the system is either a Gibbs state at a different temperature with respect to the one at which it thermalizes to, or a Generalized Gibbs state which depends on the initial data through the expectation value of some conserved quantity on that state. See, for example, Ref.~\cite{ueda2020quantum} for an overview of both theoretical and experimental aspects. In this work, we consider the simple $d$-dimensional model proposed in Ref.~\cite{saha2024prethermalization}, which shows prethermal behavior induced by a weak interaction with a spatially correlated bosonic bath, in the framework of non-equilibrium thermodynamics. Specifically, we consider the Two-Point Measurement (TPM) scheme and the End-Point Measurement (EPM) ones, which are crucial operational tools for the quantification of thermodynamically relevant quantities in stochastic and quantum thermodynamics~\cite{Campisi}. The latter, in particular, has been proposed in Ref.~\cite{gherardini2021end, hernandez2023experimental}  to characterize the energy exchange statistics and fluctuations in the presence of quantum coherence. 

Prethermalizing systems thus emerge from our study as ideal an arena for an investigation on the role of initial quantum correlations in non-equilibrium quantum thermodynamics. The prethermal phase allows for the study of the difference in the energy-exchange processes with the environment when either quantum coherences or entanglement are present in the initial state of the system. Through the study of a heat-exchange fluctuation relation resulting from the use of either the EPM or TPM protocol, we quantify the deviations due to  thermal fluctuations between the classical case and the coherence-dominated one. The latter generally results in a significant decrease in the degree of thermodynamic irreversibility, as quantified by entropy production. 

This paper is organized as follows: in Sec.~\ref{sec: pretherm} we present a brief overview of the model addressed in the remainder of this work, giving particular emphasis on the relevant physical properties. In Sec.~\ref{sec: energy} we introduce the prescriptions given by the TPM and EPM schemes for the energy statistics and apply them to a two-level system undergoing the prethermal dynamics. We discuss the role of initial energy coherences and entanglement in setting the gap between the two operational approaches. In Sec.~\ref{sec: FTs} we explore the fluctuation theorem for heat exchange for both protocols focusing on the deviations from the standard formulation given by the EPM scheme, we also discuss the dependence on the measurement basis and study the entropy dynamics. 
Finally, in Sec. \ref{sec: conclusions} we draw our conclusions. 

\section{A simple model of prethermalization}
\label{sec: pretherm}
Let us start by briefly reviewing the system under scrutiny, following the assessment presented in Ref.~\cite{saha2024prethermalization}. We consider two non-interacting qubits, each governed by the usual Zeeman Hamiltonian $H_{S,i} = {\omega_0}\sigma_z^i/2$. Here, $i=1,2$ is a label for the qubits and we have assumed units such that $\hbar = 1$. Each qubit is weakly coupled to a spatially correlated bosonic bath in thermal equilibrium at the inverse temperature $\beta$. Using the secular and Born-Markov approximation, it is shown that the resulting dynamics of the two qubits can be well approximated with the following quantum master equation
\begin{equation}
     \frac{d\rho_S}{dt} =\sum_{i=1,2} {\cal L}_{ii}(\rho_S)+\alpha\sum_{i\neq j=1,2}{\cal L}_{ij}(\rho_S)\equiv{\cal L}(\rho_S)
\label{eq:QME}
\end{equation}
with $\alpha\in]0,1[$ the spatial correlation function as defined in Ref.~\cite{saha2024prethermalization} and ${\cal L}_{ij}(\cdot)=A{\cal L}^+_{ij}(\cdot)+B{\cal L}^-_{ij}(\cdot)$. We have defined the superoperator
\begin{equation}      
    {\cal L}^+_{ij}(\cdot)=
    2\sigma_+^i\cdot\sigma_-^j - \left\{\sigma_-^j\sigma_+^i,\cdot\right\},
\label{eq:QMEop}
\end{equation}
while ${\cal L}^-_{ij}(\cdot)$ is obtained from Eq.~\eqref{eq:QMEop} by taking $\sigma_+\leftrightarrow\sigma_-$. Here, $A$ and $B$ are the bath correlation functions, related to the bath's inverse temperature $\beta$ through the relation $\beta=\frac{2}{\omega_0}\tanh^{-1}\left({\frac{B-A}{B+A}}\right)$, and we have used the Pauli ladder operators $\sigma_\pm=(\sigma_x\pm\sigma_y)/2$ with $\sigma_j$ the $j=x,y,z$ Pauli matrix. The crucial feature of Eq.~\eqref{eq:QME} is  in the dependence of the steady state solution on the parameter $\alpha$: for $\alpha < 1$ the Liouvillian ${\calL}$ has a single null eigenvalue, so that the steady-state solution is the Gibbs state $\rho_S^{ss}=e^{-\beta\sum_{i=1,2} H_{S,i}}/Z$ with $Z$ the corresponding partition function, which bears no dependence on the initial state. On the other hand, for $\alpha = 1$, ${\calL}$ two null eigenvalues appear  due to the presence of a conserved quantity, namely $\vec{\sigma_1}  \cdot \vec{\sigma_2}$. The steady-state solution has thus a dependence on the initial configuration $\rho_0$, and achieves the form of the Generalized Gibbs Ensemble (GGE)
\begin{equation}
    \rho_S^{ss} = \frac{1}{\calZ}\exp\left[-\beta\sum_{i=1,2} H_{S,i}-\frac{\ell(\rho_0)}{4}\sum_{j=x,y,z}\sigma^1_j\otimes \sigma^2_j\right] \,,
    \label{eq:GGE}
\end{equation}
with $\calZ$ the partition function of such GGE and 

where we have introduced the Lagrange multiplier $\ell(\rho_0)$ reading 
\begin{equation}
\ell(\rho_0) {=} 
\begin{cases}
0 & \text{for  } F = - \frac{3}{4},\\
\ln\left[\frac{1 - 4F}{3 + 4F}(1 +  \cosh(\beta\omega_0))\right] & \text{for  } -\frac{3}{4} {<} F {<} \frac{1}{4}, \\
\ln\left[1 + \cosh(\beta\omega_0)\right] & \text{for  } F = \frac{1}{4}.
\end{cases}
\label{eq:lagr_mult}
\end{equation}
The dependence of $\ell(\rho_0)$ on the initial state is contained in the parameter $F = \sum_{j=x,y,z}M_{jj}|_{t=0}$, which encodes the axial magnetizations $M_{jj}= \frac{1}{4}{\rm Tr}\left[(\sigma^1_j\otimes \sigma^2_j) \rho_t \right]$. 

Prethermalization has been predicted for quasi-integrable quantum many-body systems, i.e. systems that weakly violate the conservation laws specific of their integrable versions \cite{angles2020prethermalization,marcuzzi2013prethermalization,marcuzzi2016prethermalization}. These systems display a transient phase, occurring before full thermalization, during which either they reach a thermal state characterized by a temperature that is different from their respective asymptotic one or, more generally, they end up in a GGE as the one reported above. This behaviour is observed in our case study as well when the spatial correlation function $\alpha$ approaches $1$. In this case, before thermalizing at the temperature of the bosonic bath, the system enters a phase in which it is described by Eq.~\eqref{eq:GGE}. The characteristic equilibration time is proportional to $\frac{1}{|\lambda_0 - \lambda_1|}$, which corresponds to the asymptotic decay rate, where $\lambda_0$ and $\lambda_1$ are the two smallest eigenvalues of the operator $\calL$. Consequently, as $\alpha \to 1$, the spectral gap $|\lambda_0 - \lambda_1|$ decreases, leading to a longer-lived prethermal phase. 
It is worth noticing that the prethermal behaviour in this model can be witnessed already for a system as small as two qubits, thus making it very suited to simple numerical simulations. 

\section{Energy exchange in the prethermal phase}
\label{sec: energy}
We now show how the prethermal phase leaves signatures at the level of energy variations in a two-qubit system and how it can be used to probe the role played by initial quantum correlations in its thermodynamic behavior. It is well-known that the TPM scheme allows to assess energy fluctuations in a non-equilibrium process~\cite{esposito2009nonequilibrium}: through energy measurements performed at two different times, say $t=0$ and $t=\tau>0$, on a system subjected to a quantum channel $\Phi_t[\cdot]$, one can access the statistics of energy changes. However, such protocol is not apt to characterize the thermodynamics of genuinely quantum systems as it does not always reproduce the correct marginal distributions of the energy of the system at $t=t_i$ and $t=t_f>t_i$~\cite{lostaglio2022kirkwood}. To overcome this limitation, several other protocols have been proposed. Among them, the EPM approach --  which will be deployed in our study -- allows to recover the correct energy marginals, accounting for the presence of coherences in the energy eigenbasis, at the cost of introducing a non-linear dependence on the initial state in the energy statistics~\cite{gherardini2021end,hernandez2023experimental}. 

Suppose we have a $d-$dimensional quantum system that undergoes an open dynamics under the action of a given completely positive, trace-preserving (CPTP) map $\Phi_t[\cdot]$. Let $E_i$ and $\Pi_i$ ($i=1,...,d$) be the energies of the system and the projectors onto the corresponding eigenstates, respectively. The TPM and EPM protocol give the following prescriptions for the probability of measuring an energy difference $\Delta E_{if} = E_f -E_i$ at times $t_i$ and $t_f$ given that the system is prepared in state $\rho_0$ and undergoes a general time-dependent process. Specifically, calling $p^{\rm P}(\Delta E_{if})$ the probability to find the energy difference $\Delta E_{if}$ under the action of the inference protocol ${\rm P}={\TPM,\,\EPM}$, we have
\begin{equation}
   \begin{aligned}
       & p^{\TPM}(\Delta E_{if}) = {\rm Tr}[\Pi_f\Phi_{t_f}[\Pi_i\rho_0\Pi_i]\Pi_f] \,, \\
       & p^{\EPM}(\Delta E_{if}) = {\rm Tr}[\Pi_i\rho_0]{\rm Tr}[\Pi_f\Phi_{t_f}[\rho_0]] \,.
   \end{aligned}
    \label{eq:pdf}
\end{equation}
Such distributions can be used to evaluate the statistics of energy exchanges induced by the quantum channel. In particular, they allow to calculate the expectation value of the energy difference in the respective protocol $\langle\Delta E_{if}\rangle_P$. Any discrepancy between $\langle\Delta E_{if}\rangle_\TPM$ and $\langle\Delta E_{if}\rangle_\EPM$ should be ascribed to the effect of quantum coherences, which is washed out by the first energy measurement requested in the TPM protocol. In fact, defining $\Delta E_{coh}=\langle\Delta E_{if}\rangle_{\EPM} - \langle\Delta E_{if}\rangle_{\TPM}$, we have
\begin{equation}
\label{eq: prot_diff}
    \Delta E_{coh}= \sum_{i=1}^d E_i{\rm Tr}[\Pi_i\Phi_{t_f}[\chi]],
\end{equation}
where $\chi = \rho_0-\calD[\rho_0]$, $\calD[\cdot]$ being a completely dephasing map in the energy basis at time $t_i$. From this result, one can establish the  sufficient condition for the two protocols to give the same expectation value as $[\rho_0,\Pi_i]=0\,,\forall i=1,...,d$.

\begin{figure} [t!]
    \includegraphics[width=\columnwidth]{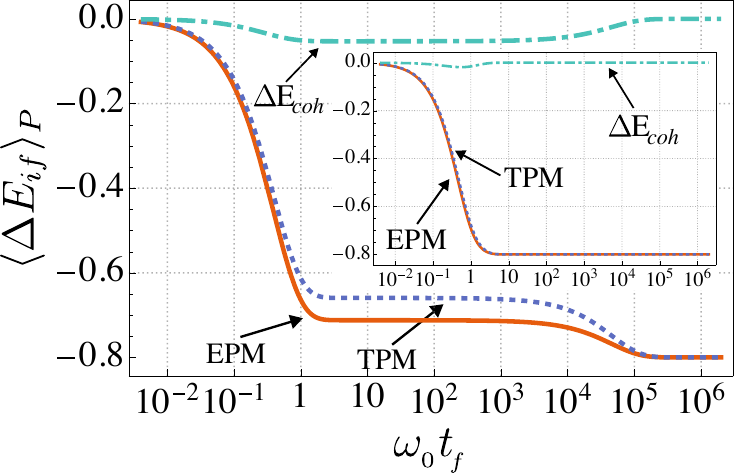}
    \caption{Mean energy difference $\langle\Delta E\rangle_P$ at growing values of the  final measurement time (in dimensionless units). We compare the results achieved through the protocols $P=\EPM,\TPM$ when choosing an initial state as in Eq.~\eqref{inistate} with $\chi = (0.2\sigma^1_x)\otimes(0.3\sigma^2_x)$. The results are typical.  
   Main panel: Prethermal dynamic with $\alpha = 1-10^{-5}$ for $A/\omega_0=0.1$ and $B/\omega_0=0.9$. We have set $t_i=0$ for convenience. We also report the coherent energy differences $\Delta E_{coh}$. Inset: Same as in the main panel but for $\alpha = 0.5$. The two protocols give different results in the prethermal phase. The differences between the two approaches for the quantification of energy exchanges that are visible in the prethermal dynamics disappear in the thermalizing phase.}
\label{fig:fluct_N2}
\end{figure}

We now apply these prescriptions to our case-study to get the mean value of the energy difference evaluated at different values of $t_f$ (setting $t_i=0$ for convenience) and measuring in the computational basis $\left\{\ket{i}\otimes\ket{j} \right\}_{i,j=0,1}$ that diagonalizes the system's Hamiltonian $H=\sum_{i=1,2}H_{S,i}$. Here, $\sigma_z\ket{0}=\ket{0}$ and $\sigma_z\ket{1}=-\ket{1}$. As in Ref.~\cite{saha2024prethermalization}, we take $A/\omega_0=0.1$ and $B/\omega_0=0.9$ for clarity of presentation of the results. In order to systematically assess the role of quantum coherences, we take an initial maximally mixed state (MMS) with coherences in the measurement basis, reading 
\begin{equation}
\label{inistate}
\rho_0 = \frac{1}{4}\openone_4 + \chi
\end{equation}
with $\openone_4$ the $4\times4$ identity matrix and matrix $\chi$ chosen to be traceless and such that $\rho_0\ge0$ while $[H,\rho]\neq0$. 
The prethermalizing phase manifests itself through the appearance of an early-time plateau in the coherent energy differences $\Delta E_{coh}$ that cannot be associated with any asymptotic properties of the system. Both the EPM and TPM protocols succeed in witnessing such plateau, as visible from the main panel in Fig.~\ref{fig:fluct_N2}, where the value of $\alpha$ has been chosen so as to instigate the prethermalization mechanism. However, the presence of quantum coherences in the initial state of the system, which are fully captured by EPM, affects quantitatively $\langle\Delta E_{if}\rangle_\TPM$, which is larger in modulus, in the prethermalizing phase, than $\langle\Delta E_{if}\rangle_\EPM$. Needless to say, as the system enters the proper thermalizing phase, both protocols deliver the same quantitative energy differences, thus bringing $\Delta E_{coh}$ to zero, as the influence of quantum coherences is depleted. For dynamics where the prethermalization process is inhibited and quantum coherences strongly suppressed by the effects of the environment, such as in the inset of Fig.~\ref{fig:fluct_N2}, the differences between EPM and TPM are virtually negligible.

\begin{figure} [h!]
    {\bf (a)}\\
    \includegraphics[width=\columnwidth]{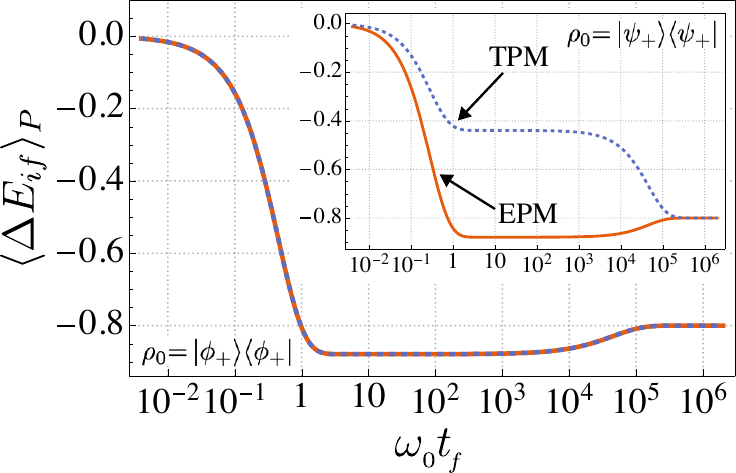}\\
    {\bf (b)}\\
    \includegraphics[width=\columnwidth]{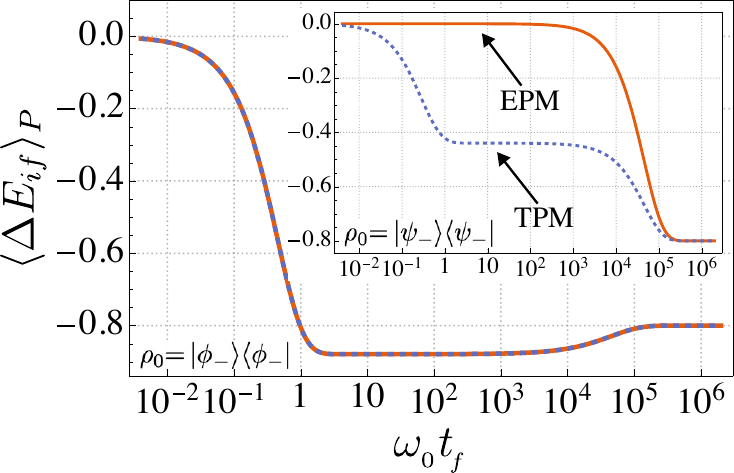}
    \caption{Mean energy difference $\langle\Delta E\rangle_P$ at growing values of the  final measurement time (in dimensionless units) when the initial state is one of the maximally entangled Bell states and in prethermal conditions given by $\alpha =1-10^{-5}$. In panel {\bf (a)} [{\bf (b)}] we show the results valid for even-parity [odd-parity] Bell states: main panel [inset] is for $\ket{\phi_+}$ [$\ket{\phi_-}$], while the inset shows the results for the preparation of $\ket{\phi_+}$ [$\ket{\phi_-}$]. Despite the maximum entanglement, showcasing strong quantum coherence, the choice of $\ket{\phi_\pm}$ does not results in a discrimination of performance between EPM and TPM during the prethermal phase, which is instead well evident for the choice of $\ket{\psi_\pm}$ as initial states.
    }
\label{fig:ent_fluct}
\end{figure}

The inspection of Eq.~\eqref{eq: prot_diff} also clarifies that the {\it nature} of the coherences shared by the subsystems of the compound that we have studied is important. In fact, in general, the mere presence of quantum correlations  is not enough to make the performance of the two protocols different. This is clearly visible in Figure \ref{fig:ent_fluct} where the previous analysis with $\alpha=1-10^{-5}$ is repeated using the elements of the maximally entangled Bell basis $\ket{\phi_\pm} = ({\ket{00}\pm\ket{11}})/{\sqrt{2}}$, $\ket{\psi_\pm} = ({\ket{01}\pm\ket{10}})/\sqrt{2}$ as initial states. In particular, the two protocols coincide for $\rho_0=\ket{\phi_\pm}\bra{\phi_\pm}$, by the mean energy variation does not overlap otherwise. Notice that, for $\alpha=1$, state $\ket{\psi_-}$ (which results in $F=-3/4$) is an eigenstate of $\mathcal{L}$ in \cref{eq:QME}. Therefore, in this case, there is no conserved quantity.

\section{Fluctuations and entropy}
\label{sec: FTs}
\subsection{The heat exchange fluctuation theorem}
We now investigate the heat exchange fluctuation theorem (XFT) for the system in the prethermal phase, highligthing how the EPM protocol leads to deviations from the standard expression.

\begin{figure} [t!]
    \centering
    \includegraphics[width=0.9\columnwidth]{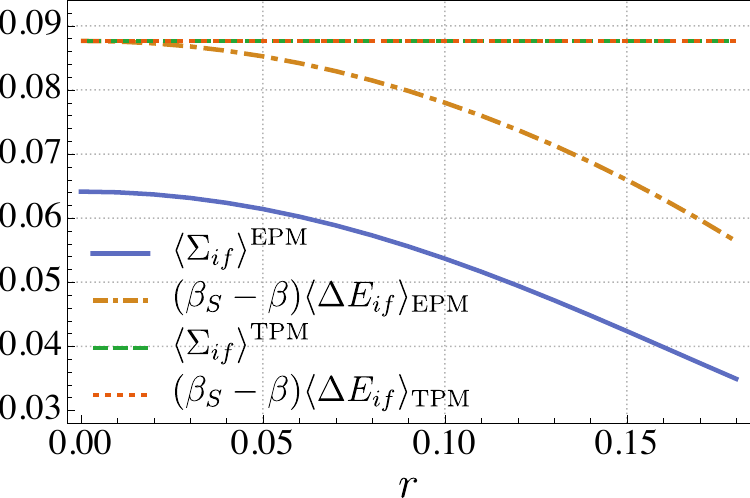}
    \caption{$\Sigma^{\EPM}$ and $\Sigma^{\TPM}$ (blue solid line and green dashed line, respectively) and their corresponding classical counterparts (orange dotdashed and red dotted lines) for a system of two identical qubits prepared in the factorized state $\rho_0=\rho^1_0\otimes\rho^2_0$ with local thermal populations and non null quantum coherences as per Eq.~\eqref{eq:init_data}. Here $a_{1,2}=r$, while $\beta_S=3\beta/2$, with $\beta\omega_0 = \ln{9}$ and $\omega_0t_f=50$. The two schemes give different thermodynamic characterizations of the same process. 
    }
\label{fig:XFTs}
\end{figure}

In the context of stochastic thermodynamics, XFTs are used to characterize the process of heat exchange between two bodies that are initially at equilibrium at two different temperatures and mutually isolated and then connected, for a time $\tau$, through a weak interaction~\cite{jarzynski2004classical}. Under reasonable assumptions, for classical systems and some restricted cases of quantum systems, using the TPM protocol, it is possible to show that
\begin{equation}
    \Sigma^{std}(Q):=\ln\left[\frac{p_{\tau}(+Q)}{p_{\tau}(-Q)}\right] = \Delta\beta \, Q
    \label{eq:clXFT}
\end{equation}

where $p_{\tau}(q)$ is the probability that the two bodies exchange an amount of heat $q$. The latter is defined as the internal energy difference of one of the two bodies (neglecting the work needed to turn on and off the interaction between them), while $\Delta \beta$ is the difference between the initial inverse temperatures of the parties at hand. Remarkably, this holds regardless of how far from equilibrium the two bodies are at time $\tau$, when the interaction is turned off. Tipically, $\Sigma^{std}$ is identified with the entropy production of the transformation \cite{jarzynski2004classical}. However, in general, such classical XFT does not hold for non-unital quantum evolutions \cite{goold2014energetic} and fails to capture true quantum effects in thermodynamics due to the use of the TPM protocol. In our context, we can study this figure of merit considering values of $\tau$ such that our system is in the prethermal phase and 
using the prescription given by the EPM protocol for the calculation of the probabilities of heat exchange. Thus, we define the entropy production as
\begin{equation}
    \!\!\Sigma^{\EPM}_{if} := \ln\left[\frac{p_{\tau}^{EPM}(\Delta E_{if})}{p_{\tau}^{EPM}(\Delta E_{fi})}\right] = \ln\left[\frac{{\rm Tr}[\rho_0\Pi_i]{\rm Tr}[\rho_S^{ss}\Pi_f]}{{\rm Tr}[\rho_0\Pi_f]{\rm Tr}[\rho_S^{ss}\Pi_i]}\right],
\end{equation}
where again $\rho_S^{ss}=\Phi_{\tau}[\rho_0]$ as prescribed by Eq.~\eqref{eq:GGE}. Consider an initial separable state of the two qubits whose local populations follow Gibbs distributions with inverse temperature $\beta_S\neq \beta$ and being endowed with local coherences, in the computational basis. That is, given $\rho^i_{\beta_S}=\exp[-\beta_s\omega_0\sigma^i_z/2]/{\cal Z}$ (with $(i=1,2)$ and ${\cal Z}={\rm Tr}[\rho^i_{\beta_S}]$) and $\chi^i = 
    \begin{bmatrix}
        0
        & a_i \\
        a_i^* & 0
        \end{bmatrix}$, the initial state is
\begin{equation}
\label{eq:init_data}
\rho_0 = (\rho^1_{\beta_S}+\chi^1) \otimes (\rho^2_{\beta_S}+\chi^2)\,.
\end{equation}
Here, we have set $a_i = r_je^{i\theta_j}$ with $|r_j|^2 < {1}/{\calZ^2}$ to guarantee $\rho_0\ge0$. For such a state, we have $F=r_1r_2\cos(\theta_1-\theta_2)+\frac{1}{4}\tanh(\frac{\beta_S\omega_0}{2})$. This results in  predictions that differ from the "standard" XFT of Eq.~\eqref{eq:clXFT} and, in order to highlight the deviations, we can write 
\begin{equation}
\begin{aligned}
    \Sigma^{\EPM}_{if} &= (\beta_S-\beta)\Delta E_{if} + \Gamma_{if},~\text{with} \\
    \Gamma_{if} &= \ln\left[e^{\beta\Delta E_{if}}\frac{{\rm Tr}[\rho_S^{ss}\Pi_f]}{{\rm Tr}[\rho_S^{ss}\Pi_i]}\right],
    \end{aligned} 
    \label{eq:XFTepm2}
\end{equation}
where $\{\Pi\}_{i=0,...,3}$ are the projectors on the computational basis. Notice that although in the above relation initial coherences only affect $\Gamma_{if}$ through $\rho_S^{ss}$, which depends on $F$, once we take the average, the term $\Delta E = \langle\Delta E_{if}\rangle$ will depend on them as well through the EPM's probability distribution. In order to show these behaviors, we consider the specific case of $\beta_S = \frac{3}{2}\beta$ (with $\beta\omega_0 = 2\ln{3}$ for the choice of  parameters made in this simulation) with $a_1=a_2$ and vary $r$ in the range $[0,{1}/{\calZ}]$ (for two identical qubits, only the modulus of the coherences matter since the only dependence on them is through $F$ as reported above. For this reason we can take them real for simplicity). We then take $\alpha = 1-10^{-5}$ and a final time that guarantees that the state is in the prethermal phase to study the averaged version of the entropy productions reported above. The results corresponding to $\omega_0t_f=50$ are reported in Fig.~\ref{fig:XFTs}, which also includes an assessment of the results stemming from the use of the TPM scheme. As one expects, the TPM protocol is insensible to the presence of initial coherences and coincides with the classical definition, despite the non-unital nature of the dynamics. This can be ascertained by checking that the "reverse" heat-exchange processes give rise to a properly normalized probability distribution $p_\tau(-q)$. Following Ref.~\cite{goold2014energetic}, this implies that the equality between the two definitions of entropy production holds on average (while being possibly broken at the level of single trajectories). On the other hand, the EPM scheme gives a coherence-dependent entropy production that deviates from the classical definition even for $r=0$, when the latter coincides with $\Sigma^{\TPM}$. This captures the different information content of the probability distributions provided by the two schemes, which in general differ even for zero initial coherences due to the classical uncertainty on the initial state present in the EPM scheme, where no initial measurement is performed~\cite{gherardini2021end}.  Overall, it is clear that the two protocols give different thermodynamic descriptions of the same process, the EPM being able to capture the effect of coherences on the entropy production and failing to satisfy the standard FT.

\subsubsection*{Basis (in)equivalence}
\begin{figure} [t!]
    \includegraphics[width=0.9\columnwidth]{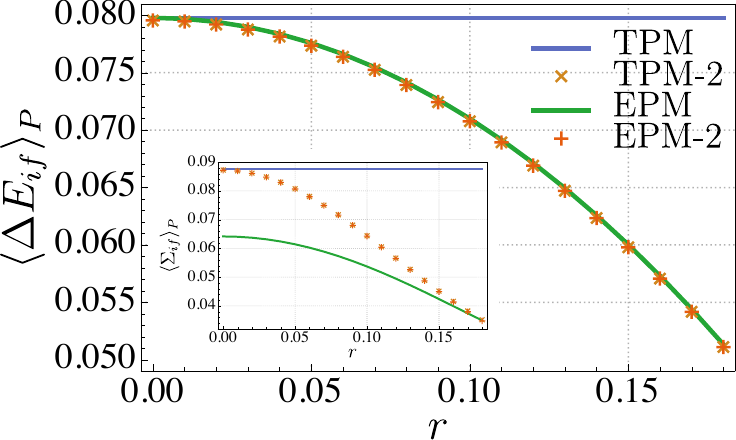}
    \caption{Main panel: Average energy against initial coherence for each of the protocols that we have addressed. Inset: Entropy production against initial coherence. We have compared the EPM and TPM schemes, measuring in the computational basis (green and blue solid lines), to the EPM-2 and TPM-2 ones with measurements in the common basis (red and orange markers). 
    }
\label{fig: common}
\end{figure}

We now explore the dependence on the choice of the energy eigenbasis, allowed by the degeneracy of the system's Hamiltonian, upon which we measure. Until now, we assumed the measurement basis to be the one that diagonalizes the system's Hamiltonian, but not the GGE. However, the  following basis of eigenstates can be sued as well $\bigl\{\ket{00}, \ket{\psi_-}, \ket{\psi_+}, \ket{11}\bigr\}$ for the two-qubits case. We bud this as ``the common basis" and we  refer to the protocols that use it as TPM-2 and EPM-2. In Fig.~\ref{fig: common} we compare the protocols using both bases by looking at the energy exchange [cf. Fig.~\ref{fig: common}{\bf (a)}] and the entropy production [cf. Fig.~\ref{fig: common}{\bf (b)}] obtained by varying the degree of initial coherence and maintaining the same setup as in the previous analysis. The energy variation in the system depends upon the choice of the measurement basis in the TPM scheme, while in the EPM scheme the results are the same, as expected. In particular, the TPM-2 agrees with both EPM and EPM-2 schemes and thus depends on the initial coherence, which is indeed referred to the computational, local basis. This should be expected: coherences in the computational basis contribute to the populations when moving to the common basis. As for the FTs, both TPM and EPM have different associated entropy productions with respect to EPM-2 and TPM-2, which in turn give the same $\Sigma$. These differences are to be expected since a different basis imply a different information extracted by the associated POVM.  

\subsection{Entropy Dynamics}
The peculiarity of the prethermal dynamics is well captured by the entropy production and its rate of change $\Pi^P=\dot{\Sigma}^P$. The Second Law of thermodynamics imposes the non-negativity of the entropy production rate, which is zero only when the system thermalizes. Thus, it is generally conceived as a measure of the system's distance from equilibrium that is useful, for example, to characterize dissipation in non-equilibrium steady states, where a constant entropy production rate is present [cf. Refs.~\cite{seifert2008stochastic,seifert2012stochastic,tome2010entropy,Imparato7a,Campisi,LandiMauro} for an in-depth analysis].

\begin{figure} [t!]
{\bf (a)}\\
    \includegraphics[width=0.9\columnwidth]{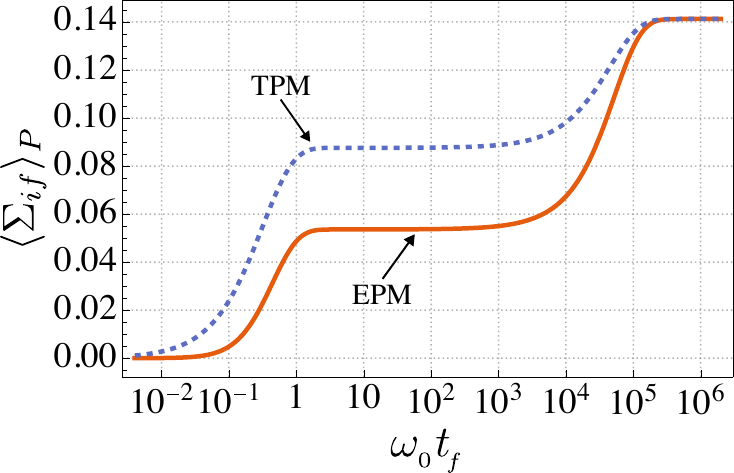}\\
   {\bf (b)}\\
    \includegraphics[width=0.9\columnwidth]{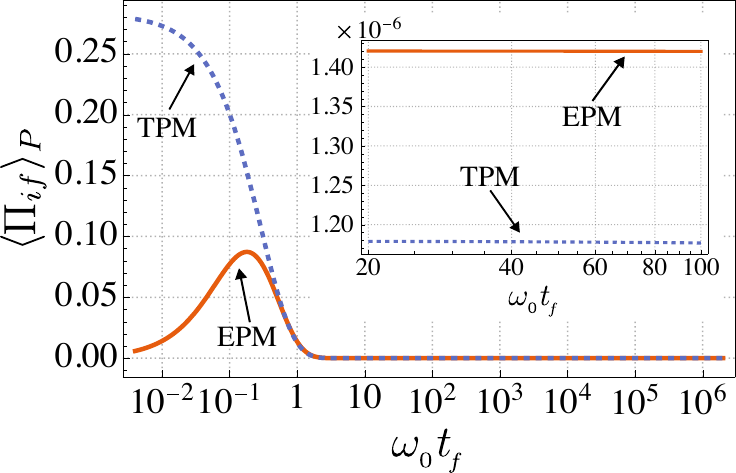}
    \caption{{\bf (a)}: Dynamics of the trajectory-averaged entropy for a system prepared in the initial state defined in \cref{eq:init_data} (identical qubits with $r=0.1$) and evaluated using both the EPM and TPM protocol. {\bf (b)}: Trajectory-averaged entropy production rate approaching the prethermal phase; In the inset we show a magnification in the region of the prethermal plateau shown in the main panel. 
    }
\label{fig: EP}
\end{figure}

Using the above definitions and taking an initial state as defined in \cref{eq:init_data}, we computed the average of the stochastic entropy production (SEP) over the whole trajectory and its time derivative, the entropy production rate, over time. The corresponding results are shown in Fig.~\ref{fig: EP}. In particular, in Fig.~\ref{fig: EP}{\bf (a)} we see that the SEP reaches the local maximum in the prethermal phase and then achieves the absolute maximum later, when the system fully thermalizes. The two protocols give a similar overall behavior, although with quantitative different values. In Figure \ref{fig: EP}{\bf (b)} we see the EP rate computed as the time derivative of the SEP in the time interval leading to the prethermal regime. The two SEP rates have different profiles, but all of them approach a null plateau. Indeed, the prethermal phase behaves in an almost indistinguishable way from a truly thermal one as far as the entropy production rate is concerned. In the indet of Fig.~\ref{fig: EP}{\bf (b)} we magnify the behavior of the SEP rate in this regime, showing how the values of $\Pi^P$ are of the order $10^{-4}$, and thus not null, which is instead the case in the thermal regime.

\section{Conclusions}
\label{sec: conclusions}
We have studied the out-of-equilibrium thermodynamics of a simple model of prethermalization using two different protocols, TPM and EPM, to highlight the role that quantum coherence have. While both protocols successfully signal the prethermal phase in the form of a plateau in the energy variation of the system, they disagree quantitatively in predicting the statistics of the energy exchanges resulting from the dynamics. This is clearly due to the existence of quantum coherences in the state of the system entering the prethermal phase, an effect that is fully captured by the EPM approach. Through the use of the heat-exchange FT and the EPM scheme, we have shown that quantum corrections to the entropy production stemming from an evolution fully within the prethermal phase are needed. Interestingly, such corrections depend on the initial quantum coherences but are non-zero even when the initial state is diagonal in the energy basis, which is due to the different information content of the state provided by the two schemes. In general, not just the amount of quantum coherences, but also their nature is important, as seen by using a measurmeent basis for the implementation o fthe EPM approach that involves entanged states. 
Our study characterizes prethermalizing dynamics as thermodynamically rich and proposes a
affirms it as a promising candidate for investigating a range of fundamental phenomena, such as the effect of initial quantum coherence on entropy production and heat exchange in genuinely nonequilibrium contexts. 

\section{Acknowledgements}
We acknowledge support from
the European Union’s Horizon Europe EIC-Pathfinder
project QuCoM (101046973), the Royal Society Wolfson Fellowship (RSWF/R3/183013), the UK EPSRC (grants EP/T028424/1 and EP/X021505/1), the Department for the Economy of Northern Ireland under the US-Ireland R\&D Partnership Programme, the ``Italian National Quantum Science and Technology Institute (NQSTI)" (PE0000023) - SPOKE 2 through project ASpEQCt.

\bibliography{bibliography}

\end{document}